\documentstyle[12pt]{article}

\begin{document}
\title{Search for flavor lepton number violation in slepton decays at LEP2
and NLC}
\author{N.V.Krasnikov \thanks{On leave of absence from INR, Moscow 117312}
\\TH Division, CERN, CH 1211, Geneva 23, Switzerland\\}

\date{November, 1995}
\maketitle
\begin{abstract}
We show that in supersymmetric models with explicit flavor lepton number
violation due to soft supersymmetry breaking mass terms there must be
flavor lepton number violation in slepton decays. We propose to look for
flavor lepton number violation in righthanded selectron and smuon decays.
For selectron and smuon lighter than 80 Gev flavor lepton number violation
in slepton decays could be discovered at LEP2 provided the mixing between
selectron and smuon is not small. We also estimate NLC discovery potential
of the lepton flavor number violation in slepton decays.
\end{abstract}
\newpage

Supersymmetric electroweak models offer the simplest solution of the
gauge hierarchy problem \cite{1}-\cite{4}. In real life supersymmetry has
to be broken and the masses of superparticles have to be lighter than
$O(1)$ Tev  \cite{4}. Supergravity gives natural explanation of the
supersymmetry breaking, namely, an account of the supergravity breaking in
hidden sector leads to soft supersymmetry breaking in observable sector
\cite{4}. For the supersymmetric extension of the Weinberg-Salam model
  soft supersymmetry breaking terms usually consist of the gaugino mass
terms, squark and slepton mass terms with the same mass at Planck scale and
trilinear soft scalar terms proportional to the superpotential \cite{4}.
For such "standard" supersymmetry breaking terms the lepton flavor number
is conserved in supersymmetric extension of Weinberg-Salam model.
However, in general, squark and slepton soft supersymmetry
breaking mass terms are not diagonal due to many reasons \cite{5}-\cite{15}
(an account of stinglike or GUT interactions, nontrivial hidden sector, ..)
and  flavor lepton number is explicitly broken due to
nondiagonal structure of slepton soft supersymmetry breaking mass terms.
As a consequence such models predict lepton flavor number violation in
$\mu$- and $\tau$-decays \cite{5}-\cite{13}. In our previous note \cite{16}
we proposed to look for flavor lepton number violation at LEP2 in
righthanded selectron and smuon decays.

In this paper we investigate the "discovery potential" of LEP2 and
NLC(next linear collider) of flavor lepton number violation in slepton decays.
We find that at LEP2 it would be possible to detect flavor lepton number
violation in selectron and smuon decays for slepton masses up to 80 Gev
provided that the mixing between selectron and smuon is not small.

In supersymmetric extensions of the Weinberg-Salam model supersymmetry is
softly broken at some high energy scale $M_{GUT}$ by generic soft terms
\begin{eqnarray}
-L_{soft} = m_{3/2}(A^{u}_{ij}\tilde{u}_R^j\tilde{q}_L^iH_u +
A^{d}_{ij}\tilde{d}^j_R\tilde{q}^i_LH_d + \nonumber\\
A^l_{ij}\tilde{e}_R\tilde{l}_LH_d +  h.c.) +
(m_q^2)_{ij}\tilde{q}_L^i(\tilde{q}^i_L)^+ + (m^2_u)_{ij}\tilde{u}^i_R
\nonumber \\
(\tilde{u}^j_R)^+ +(m^2_d)_{ij}\tilde{d}^i_R(\tilde{d}^j_R)^+ +
(m_l^2)_{ij}\tilde{l}^i_L(\tilde{l}^j_L)^+ +(m^2_e)_{ij}\tilde{e}^i_R \nonumber
\\
(\tilde{e}^j_R)^+ + m^2_1H_uH_u^+ + m_2^2H_dH_d^+ + \nonumber \\
(Bm^2_{3/2}H_uH_d +\frac{1}{2}m_a(\lambda\lambda)_{a} + h.c.) \ ,
\end{eqnarray}
where $i, j, a$ are summed over 1,2,3 and $\tilde{q}_{L}$, $\tilde{u}_{R}$,
$\tilde{d}_{R}$ denote the left- (right-)handed squarks, $\tilde{l}_{L}$,
$\tilde{e}_{R}$ the left- (right-)handed sleptons and $H_u$, $H_d$ the two
Higgs doublets; ${m}_a$ are the three gaugino masses of $SU(3)$,
$SU(2)$ and $U(1)$ respectively. In most analysis the mass terms are supposed
to be diagonal at $M_{GUT}$ scale and gaugino and trilinear mass terms are also
assumed universal at $M_{GUT}$ scale. The renormalization group
equations for soft parameters \cite{17} allow to connect high energy
scale with observable electroweak scale. The standard consequence of such
analysis is that righthanded sleptons $\tilde{e}_R$, $\tilde{\mu}_R$ and
$\tilde{\tau}_{R}$ are the lightest sparticles among squarks and sleptons.
In the approximation when we neglect lepton Yukawa coupling constants they
are degenerate in masses. An account of the electroweak symmetry breaking
gives additional contribution to righthanded slepton square mass equal
to the square mass of the corresponding lepton and besides an account
of lepton Yukawa coupling constants in the superpotential leads to the
additional contribution to righthanded slepton masses
\begin{equation}
\delta M^{2}_{sl} = O(\frac{h_{l}^{2}}{16{\pi}^2})M^{2}_{av}
ln(\frac{M_{GUT}}{M_{av}})
\end{equation}
Here $h_{l} $ is the lepton Yukawa coupling constant and $M_{av}$
is the average mass of sparticles. These effects lead to the splitting between
the righthanded slepton masses of the order of
\begin{equation}
(\delta_{e{\mu}})_{RR} =
\frac{(m^2_{\tilde{\mu}_{R}} - m^2_{\tilde{e}_R})}{m^2_{\tilde{e}_R}} =
O(10^{-5}) - O(10^{-3}) \ ,
\end{equation}
\begin{equation}
(\delta_{e{\tau}})_{RR} =
\frac{(m^2_{\tilde{\tau}_R} - m^2_{\tilde{e}_R})}{m^2_{\tilde{e}_R}} =
O(10^{-3}) - O(10^{-1})
\end{equation}
For nonzero value of trilinear parameter $A$ after electroweak symmetry
breaking we have nonzero mixing between righthanded and lefthanded sleptons,
however the lefthanded and righthanded sleptons differ in masses (lefthanded
sleptons are slightly heavier), so the mixing between righthanded and
lefthanded
sleptons (for $\tilde{e}_R$ and $\tilde{\mu}_R$) is small and we
shall neglect it. In our analysis we assume that the lightest stable particle
is gaugino corresponding to $U(1)$ gauge group that is now more or less
standard assumption \cite{18}. As it has been discussed in many papers
\cite{5} - \cite{15} in general we can expect nonzero nondiagonal soft
supersymmetry breaking terms in Lagrangian (1) that leads to additional
contributions for flavor changing neutral currents and to flavor lepton number
violation. From the nonobservation of $\mu \rightarrow e + \gamma $ decay
($Br(\mu \rightarrow  e + \gamma) \leq 5\cdot 10^{-11}$ \cite{20})  one can
find that \cite{5,6}-\cite{19}
\begin{equation}
(\delta_{e\mu})_{RR} \leq 10^{-1}M^2_{av}/(1 Tev)^2
\end{equation}
For $m_{\tilde{e}_{R}} =70 Gev$ we find that $(\delta_{e\mu})_{RR}
\leq O(10^{-3})$. Analogous bounds resulting from the nonobservation
of $\tau \rightarrow e \gamma$ and $\tau \rightarrow \mu \gamma$ decays
are not very stringent \cite{5,6}-\cite{19}.

The mass term for righthanded
$\tilde{e}_{R}$ and $\tilde{\mu}_{R}$ sleptons has the form
\begin{equation}
-\delta{L} = m_{1}^{2}\tilde{e}_{R}^{+}\tilde{e}_{R} +
m_{2}^{2}\tilde{\mu}_{R}^{+}\tilde{\mu}_{R} +
m^{2}_{12}(\tilde{e}_{R}^{+}\tilde{\mu}_{R} +
\tilde{\mu}^{+}_{R}\tilde{e}_{R})
\end{equation}

After the diagonalization of the mass term (6) we find that the eigenstates of
the mass term (6) are
\begin{equation}
\tilde{e}_{R}^{'} = \tilde{e}_{R}\cos(\phi) + \tilde{\mu}_{R}\sin({\phi}) \ ,
\end{equation}
\begin{equation}
\tilde{\mu}_{R}^{'} = \tilde{\mu}_{R}\cos(\phi) - \tilde{e}_{R}^{'}\sin(\phi)
\end{equation}
with the masses
\begin{equation}
M^{2}_{1,2} = (1/2)[(m^2_1 + m^2_2) \pm ((m^2_1 - m^2_2)^2 +
4(m^{2}_{12})^2)^{1/2}]
\end{equation}
which practically coincide for small values of $m^2_1 - m^2_2$ and
$m_{12}^2$.
Here the mixing angle $\phi$ is determined by the formula
\begin{equation}
\tan(2\phi) =2m^{2}_{12}(m^2_1 -m^2_2)^{-1}
\end{equation}
The crusial point is that even for small mixing parameter $m^{2}_{12} $
due to the smallness of the difference $m^{2}_{1} - m^2_{2}$ the mixing
angle $\phi$ is in general not small (at present state of art it is
impossible to calculate the mixing angle $\phi$ reliably).
For the most probable case when the lightest stable superparticle is
superpartner of the $U(1)$ gauge boson plus some small mixing with other
gaugino and higgsino, the sleptons $\tilde{\mu}_R$,
$\tilde{e}_R$ decay mainly into leptons $\mu_R$ and $e_R$ plus U(1) gaugino
$\lambda$. The corresponding term in the Lagrangian responsible for
sleptons decays is
\begin{equation}
L_{1} = \frac{g_{1}}{\sqrt{2}}(\bar{e}_{R}\lambda_{L}\tilde{e}_{R} +
\bar{\mu}_{R}\lambda_{L}\tilde{\mu}_{R} +h.c.),\,
\end{equation}
where $g_{1}^{2}/4\pi \approx 0.13$. For the case when mixing is absent the
decay width of the slepton into lepton and LSP is given by the formula
\begin{equation}
\Gamma = \frac{g^2_1}{32\pi}M_{sl}(1 - \frac{M^{2}_{LSP}}
{M^{2}_{sl}})^{2} ,\,
\end{equation}
where $M_{sl}$ and $M_{LSP}$ are the masses of slepton and the lightest
superparticle (U(1)-gaugino) respectively.
For the case of nonzero mixing we find that the Lagrangian (11)
in terms of slepton eigenstates reads
\begin{equation}
L_{1} = \frac{g_{1}}{\sqrt{2}}[\bar{e}_{R}\lambda_{L}
(\tilde{e}_{R}^{'}\cos(\phi) - \tilde{\mu}_{R}^{'}\sin(\phi)) +
\bar{\mu}_{R}\lambda_{L}(\tilde{\mu}^{'}_{R}\cos(\phi) +
\tilde{e}_{R}^{'}\sin(\phi)) + h.c.]
\end{equation}
At LEP2 and NLC in the neglection of slepton mixing
$\tilde{\mu}_R$ and $\tilde{\tau}_R$ sleptons pair production occurs
\cite{21} via annihilation graphs involving the photon and the $Z^{0}$ boson
and leads to the production of $\tilde{\mu}_R^+ \tilde{\mu}_R^-$ and
$\tilde{\tau}_R^+ \tilde{\tau}_R^-$ pairs. For the production of
righthanded selectrons in addition to the annihilation graphs we also have
contributions from the t-channel exchange of the neutralino \cite{21} .
In the absence of mixing the cross sections can be represented in the
form
\begin{equation}
\sigma(e^+e^- \rightarrow \tilde{\mu}_{R}^{+}\tilde{\mu}_{R}^{-}) =
kA^2 ,\,
\end{equation}
\begin{equation}
\sigma(e^+e^- \rightarrow \tilde{e}^{+}_{R}\tilde{e}^{-}_{R}) =
k(A + B)^2 ,\,
\end{equation}
where $A$ is the amplitude of s-exchange, $B$ is the amplitude of t-exchange
and $k$ is the normalization factor. The corresponding expressions for
$A$, $B$ and $k$ are contained in \cite{21}. The amplitude $B$ is determined
mainly by the exchange of the lightest gaugino and its account leads
to the increase of selectron cross section by factor $k_{in} = (4-1.5)$.
As it has been mentioned before we assume that righthanded sleptons are
the lightest visible superparticles.
So sleptons decay with 100 percent probability into leptons and LSP that
leads to accoplanar events with missing transverse momentum.
The perspectives for the detection of sleptons at LEP2 have been discussed
in refs. \cite{21}-\cite{22} in the  assumption of flavor lepton number
conservation. The main background
at LEP2 energy comes from the $W$-boson decays into charged lepton and
neutrino \cite{21}. For $\sqrt{s} = 190$ Gev the cross section of
the $W^{+}W^{-}$ production is $\sigma_{tot}(W^{+}W^{-}) \approx 26pb$.
\cite{22}. For selectrons at $\sqrt{s} = 190$ Gev, selecting events with
electron pairs with $p_{T,mis} \geq 10$ Gev and the accoplanarity angle
$\theta_{ac} \geq 34^{\circ}$ \cite{21}, the only background effects
left are from $WW \rightarrow e \nu e \nu$ and $e \nu \tau \nu$ where
$\tau \rightarrow e \nu \nu$. For instance, for $M_{\tilde{e}_R} = 85$ Gev
and $M_{LSP}=30 $ Gev one can find  that the accepted
cross section is $\sigma_{ac}= 0.17 pb$ whereas the background
cross section is $\sigma_{backgr} = 0.17 pb$ that allow to detect righthanded
selectrons at the level of $5\sigma $ for the luminosity $150 pb^{-1}$ and
at the level of $11\sigma $ for the luminosity $500 pb^{-1}$. For the detection
 of righthanded smuons we have to look for  events with two accoplanar muons
 however the cross section will be (4 - 1.5) smaller than in the selectron
case due to absence of t-channel diagram and the imposition of the cuts
analogous to the cuts for selectron case allows to detect smuons for
masses up to 80 Gev. Again here the main background comes from the $W$ decays
into muons and neutrino. The imposition of more elaborated cuts allows to
increase LEP2 righthanded smuon discovery potential up to 85 Gev on smuon
mass \cite{21,22}.

Consider now the case of nonzero mixing $\sin{\phi} \neq 0$ between
selectrons and smuons. In this case an account of t-exchange diagram
leads to the following cross sections for the slepton pair production
(compare to the formulae (14,15) ):
\begin{equation}
\sigma(e^+e^- \rightarrow \tilde{\mu}^{+}_{R}\tilde{\mu}^{-}_{R}) =
k(A + B\sin^{2}(\phi))^2 ,\,
\end{equation}
\begin{equation}
\sigma(e^+e^- \rightarrow \tilde{e}^{+}_{R}\tilde{e}^-_R) =
k(A + B\cos^{2}(\phi))^2 ,\,
\end{equation}
\begin{equation}
\sigma(e^+e^- \rightarrow \tilde{e}_{R}^{\pm}\tilde{\mu}^{\mp}_{R}) =
kB^2\cos^{2}(\phi)\sin^{2}(\phi)
\end{equation}
Due to slepton mixing we have also lepton flavor number violation
in slepton decays, namely:
\begin{equation}
\Gamma(\tilde{\mu}_{R} \rightarrow \mu + LSP) = \Gamma \cos^{2}(\phi),\,
\end{equation}
\begin{equation}
\Gamma(\tilde{\mu}_{R} \rightarrow e + LSP) = \Gamma \sin^{2}(\phi),\,
\end{equation}
\begin{equation}
\Gamma(\tilde{e}_{R} \rightarrow e + LSP) = \Gamma \cos^{2}(\phi),\,
\end{equation}
\begin{equation}
\Gamma(\tilde{e}_{R} \rightarrow \mu + LSP) = \Gamma\sin^{2}(\phi)
\end{equation}
Taking into account formulae (19-22) we find that
\begin{eqnarray}
\sigma(e^+e^-  \rightarrow e^+e^- + LSP + LSP) =  k[(A +
B\cos^{2}(\phi))^2\cos^{4}(\phi) \nonumber \\
+ (A + B\sin^{2}(\phi))^2\sin^{4}(\phi) + B^2\sin^{4}(2\phi)/8] ,\,
\end{eqnarray}
\begin{eqnarray}
\sigma(e^+e^- \rightarrow  {\mu}^+ {\mu}^{-} + LSP + LSP) = k[(A +
B\cos^{2}(\phi))^2 \sin^{4}(\phi) \nonumber \\
+ (A + B\sin^{2}(\phi))^2\cos^{4}(\phi) + B^2\sin^{4}(2\phi)/8] ,\,
\end{eqnarray}
\begin{eqnarray}
\sigma(e^+e^- \rightarrow {\mu}^{\pm} + {e}^{\mp} + LSP + LSP) =
\frac{k\sin^{2}(2\phi)}{4}[(A + B\cos^{2}(\phi))^2 \nonumber \\
+ (A + B\sin^{2}(\phi))^2 + B^2(cos^{4}(\phi) + \sin^{4}(\phi))]
\end{eqnarray}
So, as a result of nontrivial slepton mixing, we expect in general the
excess of accoplanar $e^{+}{\mu}^{-}$ and $ e^{-}{\mu}^{+}$ events compared to
the standard background which comes mainly from the leptonic $ W$-decays.
Consider at first the case of the maximal mixing $\phi = \frac{\pi}{4}$.
For this particular case we find that (here as before we neglect the
difference between righthanded selectron and righthanded smuon masses)
\begin{eqnarray}
\sigma(e^+e^- \rightarrow e^+e^- + LSP + LSP) = \sigma(e^+e^- \rightarrow
{\mu}^+{\mu}^- + LSP + LSP) = \nonumber \\
\sigma(e^+e^- \rightarrow e^{\pm}{\mu}^{\mp} +LSP + LSP) = \frac{k(A^2 + (A +
B)^2)}{4}
\end{eqnarray}
Selecting $e^{+}{\mu}^{-}$ and $e^{-}{\mu}^{+}$ events with $p_{T,mis} \geq 10$
Gev
and ${\theta}_{ac} \geq 34^{\circ}$ we find in closed analogy with
results of ref. \cite{21} that the background cross section which comes
mainly from leptonic $W$-decays for $\sqrt{s}=190$ Gev is equall to 0.34 pb and
 the accepted cross section
$\sigma (e^{+}e^{-} \rightarrow \tilde{l}_R^+ \tilde{l}_R^-
\rightarrow e^{\pm}{\mu}^{\mp} + ...)$ is equal to
(for $M_{LSP}= 20$ Gev) $0.12pb$; $0.09pb$; $0.07pb$ for the slepton masses
75Gev; 80Gev and  83Gev respectively. One can find that for such slepton masses
the flavor lepton number violation will be discovered at the level
of $9\sigma$; $7\sigma$; $5\sigma$  for the
integrated luminosity $500pb^{-1}$.
For nonmaximal mixing we have analyzed two
cases:

1. Case A - the neglection of t-channel neutralino amplitude,
   selectron cross section coincides with smuon cross
section (relatively big LSP mass).

2. Case B - the selectron cross section 3 times bigger than smuon cross
section (small LSP mass).

For these two cases we determined $5\sigma$ level bound for slepton mixing
angle which can be determined at LEP2. The results are presented in table
1. In short, we have found that for slepton masses lighter than 80 Gev
LEP2 (if it will be lucky) will discover both sleptons and flavor lepton number
violation (for the case of not small slepton mixing) in slepton decays.
For the case of maximal selectron and smuon mixing we expect equal number of
$e^{+}e^{-}$, ${\mu}^{+}{\mu}^{-}$, $e^{+}{\mu}^{-}$ and $e^{-}{\mu}^{+}$
accoplanar events unlike to the standard case (mixing is absent) when there
is excess of $e^{+}e^{-}$ and ${\mu}^{+}{\mu}^{-}$ accoplanar events over the
accoplanar  $e^{+}{\mu}^{-}$ and $e^{-}{\mu}^{+}$ events due to slepton
decays with flavor lepton number conservation.

The perspectives for the detection of sleptons at NLC (for the case of zero
slepton mixing) have been discussed in ref.\cite{23}. The standard assumption
of ref.\cite{23} is that sleptons are the LSP, therefore the  only possible
decay mode is $\tilde{l} \rightarrow l + LSP$. One possible set of selection
criteria is the following:

1. $\theta_{acop} \geq 65^{\circ}$.

2. $p_{T,mis} \geq 25$ Gev.

3. The polar angle of one of the leptons should be larger than $44^{\circ}$,
the other $26^{\circ}$.

4. $(m_{ll} - m_{Z})^2 \geq 100$ $Gev^2$.

5. $E_{l^{\pm}} \geq 150$ Gev.

For $\sqrt{s} = 500$ Gev and for integrated luminosity 20 $fb^{-1}$ , a
$5\sigma$ signal can be found up to 225 Gev provided the difference between
lepton and LSP is greater than 25 Gev \cite{23}. Following ref.\cite{23} we
have analyzed the perspective for the detection of nonzero slepton mixing at
NLC. In short, we have found that for $M_{LSP}=100$ Gev it is possible to
discover selectron-smuon mixing at the $5\sigma$ level for $M_{sl}=150 $
Gev provided that $\sin{2\phi} \geq 0.28$. For $M_{sl}= 200$ Gev it is
possible to detect mixing for $\sin{2\phi} \geq 0.44$ and  $M_{sl}=225$
Gev corresponds to the limiting case of maximal mixing ($\sin{2\phi} =1$)
discovery.

It should be noted that we restricted ourselves to the case
of smuon selectron mixing and have neglected stau mixing with selectron and
smuon. In general  case the situation will be slightly more complicated.
For instance, for the case of stau-smuon mixing in formulae
(23-25) we have to put $B = 0$ ( only s-exchange graphs contribute to the
cross sections) and in final states we expect as a result of mixing
${\tau}^{\pm}{\mu}^{\mp}$ accoplanar pairs. The best way to detect
$\tau$ lepton is through hadronic final states, since
$Br(\tau \rightarrow hadrons + {\nu}_{\tau})$ = 0.74. Again, in this case the
main background comes from W-decays into $(\tau)^{\pm}(\mu)^{\mp} + \nu + \nu$
in the reaction $e^+e^- \rightarrow W^+W^-$. The imposition of some cuts
\cite{21,22} decreases W-background to $0.07pb$ that allows to detect
stau-smuon mixing for slepton masses up to 70 Gev. We have found that
for $m_{sl} = 50 $ Gev it would be possible to detect mixing angle
$\sin(2{\phi}_{\tau\mu})$ bigger than 0.70. Other detectable consequence of
big stau-smuon mixing is the decrease of accoplanar ${\mu}^+ {\mu}^- $
events compared to the case of zero mixing. For instance, for the case of
maximal mixing $\sin(2{\phi}_{\tau\mu}) = 1$ the suppression factor is 2.
In general we can't exclude big mixing between all three
righthanded sleptons and as a typical consequence of such mixing we expect
the excess of nondiagonal accoplanar lepton pairs.

 For the $e^+e^-$ energy less than 160 Gev
the WW-background is practically zero \cite{21} so for slepton masses
lighter than 70 Gev the best way to detect slepton mixing is the decrease
of the energy for LEP2 (LEP1.5).

Let us formulate the main result of this paper: in supersymmetric extension of
standard Weinberg-Salam model there could be soft supersymmetry breaking terms
 responsible for flavor lepton number violation and slepton mixing. If sleptons
are relatively light and mixing is not small it would be possible to discover
both sleptons and lepton flavor number violation in slepton decays at LEP2.

I thank CERN TH Department for the hospitality during my stay at CERN where
this paper has been finished. I am indebted to the collaborators of the INR
theoretical department
for discussions and critical comments. The research described in this
publication was made possible in part by Grant N6G000 from the International
Science Foundation and by Grant 94-02-04474-a of the Russian Scientific
Foundation.

\newpage

Table 1. LEP2 $5\sigma$ discovery potential for selectron-smuon mixing angle
$\sin{2\phi}$ for different slepton masses and for $L = 500pb^{-1}$

\begin{center}
\begin{tabular}{|l||l||l|}
\hline
$M_{sl}$ in Gev & $\sin{2\phi}$ & $\sin{2\phi}$ \\
\hline
                & case A     & case B  \\
\hline
50       & 0.60 & 0.40  \\
\hline
55       & 0.64 & 0.44 \\
\hline
60       & 0.69 & 0.46 \\
\hline
65       & 0.75 & 0.50 \\
\hline
70       & 0.85 & 0.57 \\
\hline
75       & 0.96 & 0.67 \\
\hline
80       & - & 1 \\
\hline
\end{tabular}
\end{center}

\end{document}